\documentclass{article}

\def\dz{\dot{z}}
\def\dbz{\dot{\bar z}}
\newcommand{\nn}{\nonumber\\}
\newcommand{\p}[1]{(\ref{#1})}
\newcommand{\cZ}{{\cal Z}}

\newcommand{\bH}{{\overline H}}
\newcommand{\bz}{{\bar z}}

\newcommand{\bn}{{\overline\nabla}}

\newcommand{\cbZ}{\overline{\cal Z}}

\newcommand{\cF}{{\cal F}}

\newcommand{\cbF}{\overline{\cal F}}
\newcommand{\bF}{{\overline {\rule{0pt}{0.7em} F}}{}}

\newcommand{\bxi}{{\bar\xi}}
\newcommand{\bpsi}{{\bar\psi}}
\newcommand{\disty}{\displaystyle}

\newcommand{\ba}{\begin{array}}
\newcommand{\ea}{\end{array}}
\newcommand{\be}{\begin{equation}}
\newcommand{\ee}{\end{equation}}
\newcommand{\bea}{\begin{eqnarray}}
\newcommand{\eea}{\end{eqnarray}}
\newcommand{\bi}{\begin{itemize}}
\newcommand{\ei}{\end{itemize}}

\newcommand {\bD}{\overline{D}}

\newcommand{\rea}{{\tt Re\,}A}
\newcommand{\reb}{{\tt Re\,}B}
\newcommand{\ima}{{\tt Im\,}A}
\newcommand{\imb}{{\tt Im\,}B}
\newcommand{\dima}{{\tt Im\,}\dot A}
\newcommand{\dimb}{{\tt Im\,}\dot B}
\newcommand{\M}{\mbox{M}}
\newcommand{\N}{\mbox{N}}
\newcommand{\dx}{\dot{x}}
\newcommand{\bx}{\bar x}
\newcommand{\dbx}{\dot{\bx}}

\usepackage{amscd,amsmath,amssymb}

\begin{document}
\thispagestyle{empty}
\vspace{2cm}

\begin{center}
{~}\\
\vspace{3cm}
{\Large\bf Hyper-K\"ahler geometry and dualization}\\
\vspace{2cm}
{\large \bf S.~Bellucci${}^{a}$,  S.~Krivonos${}^{b}$ and A.~Shcherbakov${}^{b}$  }\\
\vspace{2cm}
{\it ${}^a$INFN-Laboratori Nazionali di Frascati, Via E. Fermi 40,
00044 Frascati, Italy}\\
{\tt bellucci@lnf.infn.it} \\\vspace{0.5cm}
{\it ${}^b$ Bogoliubov  Laboratory of Theoretical Physics, JINR,
141980 Dubna, Russia}\\
{\tt krivonos, shcherb@theor.jinr.ru} \\ \vspace{2.5cm}
\end{center}

\begin{abstract}
We demonstrate that in $N=8$ supersymmetric mechanics with linear and nonlinear chiral supermultiplets one may dualize
two auxiliary fields into physical ones in such a way that the bosonic manifold will be a hyper-K\"ahler one. The key
point of our construction is about different dualizations of the two auxiliary components. One of them is turned into a physical
one in the standard way through its replacement by the total time derivative of some physical field. The other
auxiliary field which obeys the condition $\partial_t (\ima+\imb)=0$ is dualized through a Lagrange multiplier. We clarify
this choice of dualization by presenting the analogy with a three-dimensional case.
\end{abstract}
\newpage
\section{Introduction}
One of the most interesting features of supersymmetric sigma-models is the interplay between the
number of supersymmetries and the geometry of the bosonic sector. The following well known results 
concern $D=4$ sigma models: the geometry of the bosonic sector should be K\"ahler in the $N=1$ case \cite{D41} and
hyper-K\"ahler for $N=2$ (rigid) supersymmetry \cite{D42}. The independent consideration of $D=1$ sigma models
reveals other bosonic target geometries: HKT (hyper-K\"ahler with torsion) for $N=4$ supersymmetric theories with four
physical bosons and OKT (octonionic-K\"{a}hler with torsion) for $N=8$ ones in the case of eight physical bosonic fields
\cite{GPS}. Moreover, the detailed analysis of the general components and superfield actions for $N=4,8$ cases with
different numbers of physical bosonic degrees of freedom shows that only conformally flat  geometries with additional
restrictions on the metrics of bosonic manifolds may arise \cite{N48}-\cite{N48d}.
Of course, one may perform the
direct reduction in terms of the components from, say $N=2, D=4$ sigma model to $D=1$ and obtain a hyper-K\"ahler sigma
model with $N=8$ supersymmetry, but the question still persists: what is wrong (incomplete) with the standard components
and superfields approaches in $D=1$? Or, in other words, why do only conformally flat metrics show up? In
\cite{{BuKS},{KS},{DIK}} this question has been answered for $N=4,D=1$ sigma-models, proving that only essentially
nonlinear supermultiplets give rise to bosonic geometries, different from conformally flat ones. Roughly speaking,
the superfield constraints defining the $N=4, D=1$ off-shell supermultiplets contain arbitrary functions which play
an essential role and enter the metrics of bosonic manifolds. With such constraints the simplest ``free'' superfield
Lagrangian describes the hyper-K\"ahler sigma model in its bosonic target space. Clearly enough, more general
Lagrangians would exhibit a more general geometry in the bosonic sector, generalizing the hyper-K\"ahler one.

This situation, which seems to be more or less natural in the case of $N=4, D=1$ supersymmetry looks a little bit
puzzling when applied to the case of $N=8$ supersymmetry in $D=1$.\footnote{
For pedagogical introductions to this subject the reader is invited to consult e.g. \cite{lectures}.}
Indeed, $N=4, D=1$ supersymmetry corresponds to $N=1$
in $D=4$ and therefore the natural constraints on the bosonic sigma models are K\"ahler ones. The appearance of
hyper-K\"ahler geometries in $N=4, D=1$ sigma models is related to further additional constraints on the theory
(``free'' superfield Lagrangians). When we turn to the $N=8$ case the situation changes drastically: supersymmetry
imposes much stronger constraints on the metrics of the bosonic manifolds and no apparent reasons exist to expect an
additional functional freedom in the $N=8$ supermultiplets.  So, how may supersymmetric sigma models with non-conformally
flat metrics arise in a such situation?  In the present paper we will partially answer this question.
We will consider $N=8$ linear and nonlinear $({\bf 2,8,6})$ supermultiplets\footnote{We use the notation ({\bf
n,N,N-n}) to describe a supermultiplet with $n$ physical bosons, $N$ fermions and $N-n$ auxiliary bosons.} \cite{ABC}
and demonstrate that after proper dualization of two auxiliary bosonic components into physical ones, the resulting
bosonic sigma model is of the hyper-K\"ahler type. Performing dualization we essentially use the equations
of motion and constraints for auxiliary components to turn them into physical ones. As a result, no new functions appear
in the action besides the metric of the bosonic manifold. Our approach is reminiscent of duality transformations between a gauge
vector field and a scalar in $D=3$. In order to simplify the presentation, almost everywhere in this paper we neglect
fermionic terms which may be easily restored, if needed.

\setcounter{equation}0
\section{N=8 linear chiral supermultiplet}
We start with the linear case and show how to dualize the $N=8$ linear chiral multiplet (LCM) into a new one
which possesses a hyper-K\"ahler sigma model in its bosonic sector.

The simplest way to describe the $N=8$ LCM is to introduce a complex bosonic superfield $\cZ$ obeying
the constraints \cite{ABC}
\bea
&& D^a\cZ = 0 \; , \quad \nabla^\alpha \cZ = 0 , \label{con1l} \\
&& \bn{}^\alpha \bD{}^a \cZ + \nabla^\alpha D^a \cbZ = 0. \label{con2l}
\eea
Here, the covariant spinor derivatives $D^a, \bD_a, \nabla^\alpha, \bn_\alpha $ are defined in $N=8, d=1$
superspace $\mathbb{R}^{(1|8)}$
\be\label{ss}
\mathbb{R}^{(1|8)} = (t, \,\theta_a,\, \bar\theta^b\; ,\vartheta_\alpha ,\, \bar\vartheta{}^\beta ) ,\quad \overline{
\left( \theta_a\right)} =\bar\theta{}^a,\; \overline{\left( \vartheta_\alpha \right)} =\bar\vartheta{}^\alpha, \quad a,
b, \alpha, \beta = 1, 2
\ee
by
\bea\label{sderiv}
&& D^{a}=\frac{\partial}{\partial\theta_{a}}+i\bar\theta{}^{a}\partial_t\,,\;
\bD_{a}=\frac{\partial}{\partial\bar\theta{}^{a}}+i\theta_{a}\partial_t\,,\quad \{D^a, \bD_b\} = 2i \delta^a_b
\partial_t , \nn &&
\nabla^{\alpha}=\frac{\partial}{\partial\vartheta_{\alpha}}+i\bar\vartheta{}^{\alpha}\partial_t\,,\;
\bn_{\alpha}=\frac{\partial}{\partial\bar\vartheta{}^{\alpha}}+i\vartheta_{\alpha}\partial_t\,,\quad \{ \nabla^\alpha,
\bn_\beta \} = 2i \delta^\alpha_\beta \partial_t .
\eea

Being (anti)chiral, the superfields $\cZ, \cbZ$ contain 16 bosonic and 16 fermionic components.
Let us define these components as follows\footnote{All implicit summations go from ``up-left'' to
``down-right'', e.g., $\psi\bpsi \equiv \psi^i\bpsi_i$, $\psi^2 \equiv \psi^i\psi_i$,  etc.}:
\be\label{compl}
\ba{ll}
\mbox{Bosonic components}&\\[2mm]
 z=\cZ, & \bz=\cbZ ,\\
 A=-i\bD{}^2\cZ ,& \bar{A}=-iD^2\cbZ ,\\
 B=-i\bn{}^2\cZ ,& \bar{B}=-i\nabla^2\cbZ,\\
 Y^{a\alpha}=\bD{}^{a}\bn{}^{\alpha}\cZ ,& \bar{Y}{}^{a\alpha}=-D^{a}\nabla^{\alpha}\cbZ ,\\
 X=\bD{}^2\bn{}^2\cZ ,&\bar X = D^2 \nabla^2\cbZ
\ea\qquad \ba{ll}
\mbox{Fermionic components}&\\[2mm]
 \psi_a= \bD_a\cZ ,& \bpsi_a=-D_a\cbZ,\\
 \xi_\alpha=\bn_\alpha \cZ ,& \bxi_\alpha=-\nabla_\alpha\cbZ ,\\
\tau_\alpha = \bD{}^2 \bn_\alpha \cZ ,& \bar\tau_\alpha= D^2\nabla_\alpha\cbZ ,\\
 \sigma_a=\bn{}^2 \bD_a\cZ ,& \bar\sigma_a=\nabla^2 D_a\cbZ ,\\
 \\
\ea
\ee
where the right hand side of each expression is supposed to be taken with $\theta=\vartheta=0$.

Now it is time to consider the constraints \p{con2l}. Besides the evident reality condition on the $Y^{a\alpha}$
\be\label{Yl}
Y^{a\alpha}=\bar{Y}{}^{a\alpha},
\ee
they put the following restriction on the higher components of the superfield $\cZ$:
\bea
&& X=16\ddot\bz ,  \qquad \tau^\alpha= -4i\dot\bxi{}^\alpha, \qquad
\sigma^a= -4i\dot\bpsi{}^a,  \label{bos1l} \\
&& \frac{\partial}{\partial t} \left(\bar{A} -B\right) =0. \label{bosl2}
\eea
The equations  \p{bos1l} express the higher bosonic components $X, \bar X$ and fermionic ones $\tau^\alpha,
\bar\tau{}^\alpha, \sigma^\alpha, \bar\sigma{}^\alpha$ in terms of physical bosons and fermions.
For the auxiliary bosons  $A, B$ we have the differential equation \p{bosl2} which reduces the number of
independent auxiliary fields to six, and therefore our linear supermultiplet is just the $(2,8,6)$ one.

The most general $N=8$ sigma-model type  action in $N=8$ superspace reads \cite{{bkn},{bks1}} \footnote{We use
the convention $\int d^2 \theta = \frac {1}{4} D^{i} D_i$, $\int d^2 \bar\theta = \frac {1}{4} \bD_i \bD^i$.}
\be\label{actionl}
S =  \int\! dt d^2 \bar\theta d^2 \bar\vartheta\; \cF(\cZ) + \int\! dt
    d^2\theta d^2 \vartheta\; \cbF (\cbZ) \;.
\ee
Here  $\cF (\cZ)$ and $\cbF(\cbZ)$ are arbitrary holomorphic functions depending only on $\cZ$ and $\cbZ$,
respectively. Now we may go to components. In what follows we will be interested only in the bosonic sector of our
theory. So, discarding all fermions we will finish with the following component action:
\be\label{action1l}
S_{bos} =-\,\frac1{16} \int dt \left[ \rule{0pt}{1em} F' X + \bF' \bar X - F'' \left( 2 Y^{a\alpha} Y_{a\alpha} + A
B\right)
    - \bF'' \left( 2 {\bar Y}^{a\alpha} {\bar Y}_{a\alpha} + \bar A \bar B\right) \right].
\ee
On-shell the component $Y^{a\alpha}$ turns out to be zero in the bosonic limit, so we are left with the following action:
\be\label{action2l}
S_{bos} = \frac1{16}\int dt \left[ \rule{0pt}{1em}  F'' A B + \bF'' \bar A \bar B - F' X - \bF' \bar X  \right]
\ee
where the components $A$ and $B$ satisfy \p{bosl2} while the higher components $X, {\bar X}$ are defined in \p{bos1l}.

Now one should decide what to do with the constraints \p{bosl2}. The simplest possibility is to solve these
constraints as \cite{{bkn},{bks1}}
$$
A={\bar B}.
$$
In this case the supermultiplet will be indeed $({\bf 2,8,6})$. Another possibility is to plug these constraints into the
action \p{action2l} using Lagrange multipliers. Varying over the auxiliary fields we can express them through time
derivatives of the Lagrangian multipliers. Thus, the corresponding theory contains four physical bosons ($\cZ,
\cbZ$ and a complex Lagrange multiplier). One may check that the resulting action describes a conformally flat
four-dimensional bosonic sigma-model \cite{{bks1},{bksu}}.

It is quite unexpected that there is a third way to deal with the constraints \p{bosl2}.  First, we rewrite them as
follows:
\bea
&& \frac{\partial}{\partial t} \left( \rea-\reb \right)=0, \label{rel} \\
&& \frac{\partial}{\partial t} \left( \ima+\imb \right)=0. \label{iml}
\eea
Then, one may check that under $N=8$ supersymmetry $\rea$ and $\reb$ transform in the same way
\be\label{trl}
\delta \rea = \delta \reb = -2 \frac{\partial}{\partial t} \left(
 \epsilon^a \psi_a -  \bar\epsilon{}^a \bar\psi_a -  \bar\varepsilon{}^\alpha \bar\xi_\alpha +
 \varepsilon^\alpha \xi_\alpha \right)
\ee
so that we can solve the constraint \p{rel} by dualizing $\rea$ and $\reb$ into the same physical field
\be
\rea = \reb =4 \dot\Phi, \qquad \delta\Phi =-\, \frac12 \left(
 \epsilon^a \psi_a -  \bar\epsilon{}^a \bar\psi_a -  \bar\varepsilon{}^\alpha \bar\xi_\alpha +
 \varepsilon^\alpha \xi_\alpha \right).
\ee
Finally, we insert the second constraint \p{iml} into the action \p{action2l} with the Lagrange multiplier $y$. With this
identification the bosonic action acquires the following form:
\bea\label{actf}
&& S = \int d t \left[\rule{0pt}{1em} (F'' + \bF'') (\dot\Phi^2 + \dz \dbz) - \frac1{16} (F'' + \bF'') \ima \, \imb
    + \frac i4 \, (F'' - \bF'') \dot\Phi (\ima + \imb)  \right. \nn
    && \phantom{S=\int dt }\left. \rule{0pt}{1em} +\frac14 \,y \, \partial_t (\ima + \imb) \right].
\eea
Eliminating the fields $\ima$, $\imb$ via their equations of motion allows us to
represent the action in the Gibbons--Hawking form with two isometries
\bea\label{finall}
&S = \displaystyle  \int d t \left[ \rule{0pt}{1em}g \dot{\vec{\Phi}}^2 + \frac 1g (\dot y + \vec{A} \dot{\vec{\Phi}})^2 \right],&\nn
&g \equiv F'' + \bF'', \quad \vec\Phi = (z,\bz,\Phi), \quad \vec A = (0,0,-i(F''-\bF'')).&
\eea
The metric $g$ and the vector $\vec A$ satisfy
$$\triangle g = 0, \qquad  \vec\nabla \times \vec{A} = \vec\nabla g. $$

Thus, we see that the action \p{finall} describes a four-dimensional hyper-K\"ahler sigma model with $N=8$ supersymmetry
(provided we restore the fermionic terms). The key step in our construction is a different treatment of the constraints
\p{rel} and \p{iml}. One may wonder why we have to dualize our auxiliary components into physical ones in this way?
The proper explanation has a three-dimensional origin. In $D=3$ the analog of the constraints  \p{rel} and \p{iml} has
the following form:
\bea
&& \partial_{ac}\rea^c_b+\partial_{bc}\rea^c_a =0, \label{d31} \\
&& \partial_{ab} \ima ^{ab} =0 . \label{d32}
\eea
The constraint \p{d31} is a Bianchi identity, allowing one to express $\rea_{ab}$ in terms of a new scalar field $\Phi:
\rea_{ab}=\partial_{ab} \Phi$, whereas the constraint \p{d32} is a Bianchi identity, establishing that $\ima_{ab}$ is a field
strength. Clearly enough, in order to have two scalars, one should solve the first constraint and dualize the second one. This
is just the way we treated these constraints here. Instead, if we were to dualize both constraints, we would have, once
again, a field strength and a scalar, rather than the two additional scalars which we need for the hyper-K\"ahler geometry.

\setcounter{equation}0
\section{N=8 nonlinear chiral supermultiplet}
The $N=8$ nonlinear chiral supermultiplet (NCM) was introduced in \cite{bbks}.
This supermultiplet, similarly to the LCM
one, has a natural description in $N=8, d=1$ superspace \p{ss}. We again consider a complex bosonic superfield $\cZ$,
but now we impose the modified constraints
\bea
&& D^a\cZ = -\cZ \bD{}^a \cZ \; , \quad \nabla^\alpha \cZ = -\cZ \bn{}^\alpha \cZ , \label{con1nl} \\
&& \bn{}^\alpha \bD{}^a \cZ + \nabla^\alpha D^a \cbZ = 0. \label{con2nl}
\eea
We may see now that the NCM obeys a nonlinear deformation of the chirality conditions \p{con1nl}, while the reality
constraint
\p{con2nl} is the same as in the linear case.

Usually, when dealing with nonlinear supermultiplets it appears a problem of construction for the proper action. However,
for the NCM the most general sigma-model type action has the same form \p{actionl}. One may check that despite the
nonlinearity of the constraints, the action \p{actionl} is still invariant with respect to the full $N=8$ supersymmetry
\cite{bbks}.

The component structure of the $N=8$ superfield $\cZ$, implied by \p{con1nl}, \p{con2nl}, is a bit involved, in
comparison with the $N=8$ LCM case. In order to define it, let us firstly consider the constraints \p{con1nl}. In
contrast to the LCM case, now the derivatives $D^a$ and $\nabla^\alpha$ of the superfield $\cZ$ (or $\bD{}^a,
\bn{}^\alpha$ of $\cbZ$) are not equal to zero. Nevertheless, they do not generate new independent components because,
as it immediately follows from \p{con1nl}, they  can be expressed as $\bD{}^a$, $\bn{}^\alpha$ (or $D^a$,
$\nabla^\alpha$) derivatives of the same superfield. Therefore, as in the case of the LCM, only the components appearing in
the $\bar\theta{}^a,
\bar\vartheta{}^\alpha$ expansion of $\cZ$ and the $\theta{}^a, \vartheta{}^\alpha$ expansion of the $\cbZ$ superfield
are independent. Consequently, as a solution to the constraints \p{con1nl} we have 16 bosonic and 16 fermionic
components in the NCM which we define as previously \p{compl}.

The role of the constraints \p{con2nl} is to further reduce the number of independent components in the NCM.
One may check that the superfield constraints \p{con2nl} encompass the following component ones:
\bea\label{conNL}
Y^{a\alpha}&=&\bar{Y}{}^{a\alpha}, \nn
X&=&16\ddot\bz - 4 \partial_t(\bz\bar{A} + \bz\bar{B} -i \bpsi{}^2-i\bxi{}^2) +\bz \left( 4 Y^{a\alpha}Y_{a\alpha}
+ 2\bar{A}\bar{B}  -
 \bz \bar X\right) \nn
&& -2i \left[ \bar{B}\bpsi{}^2 +\bar{A} \bxi{}^2+4i\bxi{}^\alpha\bpsi^a Y_{a\alpha}+
     2i\bz(\bpsi{}^a\bar\sigma_a+\bxi{}^\alpha\bar\tau_\alpha )\right], \nn
4\dot{\bar{A}}-4\dot B &=&  \left[ \bar{A}\bar{B}+2(\bpsi{}^a\bar{\sigma}_a+\bxi^\alpha\bar\tau_\alpha) -\bz\bar X\right]
        - \left[ A B- 2(\psi^a\sigma_a+ \xi^\alpha\tau_\alpha)-z X\right],\nn
\tau^\alpha&=& -4i\dot\bxi{}^\alpha +\left( 2Y^{a\alpha}\bpsi_a-\bar\tau{}^\alpha \bz +i\bar{A}\bxi{}^\alpha\right),\nn
\sigma^a&=& -4i\dot\bpsi{}^a -\left( 2Y^{a\alpha}\bxi_\alpha+\bar\sigma{}^a\bz-i\bar{B}\bpsi{}^a\right).
\eea
Thus we see that $Y^{a\alpha}$ is still real, the components $X, \tau^\alpha,\sigma^a$ are expressed in terms of others, and the
complex bosonic components $A$ and $B$ obey some nonlinear differential constraints. Thus now, before any dualization, our
NCM is a $({\bf 2,8,6})$ supermultiplet.

What is really important is that the transformation properties of $\rea$ and $\reb$ are {\it exactly the same} as in
the linear case
\be\label{trNL}
\delta \rea = \delta \reb = -2 \frac{\partial}{\partial t} \left(
 \epsilon^a \psi_a -  \bar\epsilon{}^a \bar\psi_a -  \bar\varepsilon{}^\alpha \bar\xi_\alpha +
 \varepsilon^\alpha \xi_\alpha \right),
\ee
while the imaginary parts of $A$ and $B$ get nonlinearities in their transformation laws.

For simplicity of the presentation, we will discard, once again, all fermionic terms and consider only the bosonic sector of the theory. The
bosonic part of the action \p{actionl} has the same form \p{action1l}. On-shell the component $Y^{a\alpha}$ turns
out to be still zero in bosonic limit, and we are left with the action
\be\label{action2nl}
S_{bos} = \frac1{16}\,\int dt \left[ \rule{0pt}{1em} F'' A B + \bF'' \bar A \bar B - F' X - \bF' \bar X \right],
\ee
where the components $A$ and $B$ satisfy
\be\label{AB}
\ba{ll}
\partial_t \left( \rea - \reb\right) = 0, & \qquad (a) \\
4i \partial_t \left( \ima+\imb \right) = A B - \bar A \bar B - z X + \bz \bar X.& \qquad (b)
\ea
\ee
while the higher component $X$ is defined in \p{conNL}.

Our strategy is to dualize the auxiliary fields into physical ones.
The first step is evident: owing to (\ref{AB}a) and \p{trNL}, one may turn the real parts of $A$ and $B$ into a physical boson
\be\label{re_dual}
\rea = \reb = \dot\Phi.
\ee
Having this identification, one may rewrite the constraint (\ref{AB}b) as
\be\label{Bianchi}
\frac{d}{dt} \left[e^{-\textstyle\frac{\Phi}2} \left(\rule{0pt}{1em} (1 + z \bz) (\ima + \imb)
    - 4i (z \dbz - \bz \dz )\right) \right] = 0.
\ee
In full analogy with the LCM case, one should incorporate this equation in the action \p{action2nl}
\bea\label{actionNL1}
&& S_{bos} = \int dt \left\{ \rule{0pt}{1.2em} \frac1{16} \, F'' A B + \frac1{16} \, \bF''  \bar A \bar B
    - \frac1{16} \,F' X - \frac1{16} \, \bF' \bar X -
\right.\nn
&&\phantom{S_{bos} = int dt}    \left. \rule{0pt}{1.2em}\frac y4 \partial_t\left[
    e^{-{\Phi}/2} \left( \rule{0pt}{1em}(1 + z \bz) (\ima + \imb)
    - 4i (z \dbz - \bz \dz )\right)\right] \right\},
\eea
where the highest component $X$ of the superfield $\cZ$ is defined from \p{conNL} as
\bea\label{Xnl}
(1-z^2 \bz^2)X &=& 16 (\ddot\bz - \bz^2 \ddot z) + 2 \bz (1 - z\bz) \dot\Phi^2 - 8 \bz \ddot\Phi (1 - z\bz)
    - 8 \dot\Phi (\dbz - \bz^2 \dz) - 2 \bz (1 - z\bz) \ima \imb \nn
&&    + 4i \bz (1 + z \bz) (\dima + \dimb)
    - 2i (\ima + \imb) \left(\bz (1 + z \bz) \dot\Phi - 2 (\dbz + \bz^2 \dz)\right).
\eea
Before substituting $X, \bar X$ in the action \p{actionNL1}, it is useful to introduce new functions $H$ and $\bH$ defined
as
\be\label{H}
H (z,\bz) = \frac{F' - z^2 \bF'}{1 - z^2 \bz^2}, \qquad \bH (z,\bz) = \frac{\bF' - \bz^2 F'}{1 - z^2 \bz^2}.
\ee
By definition, they satisfy
\be\label{Hprop}
H_{,\bz} = - z^2 \bH_{,\bz}, \qquad \bH_{,z} = - \bz^2 H_{,z}.
\ee
Finally, combining all this together, we get the following Lagrangian:
\be\label{Lagr1}
\ba{l}
\disty L = \frac1{16} \, (1 - z^2 \bz^2) (H_{,z} + \bH_{,\bz}) \left( \dot\Phi^2 - \ima \imb \right)
    +\frac i{16} \, (1 - z^2 \bz^2) (H_{,z} - \bH_{,\bz}) \dot\Phi \left( \ima + \imb \right) + \\ [3pt]
\disty \phantom{L = }    (H_{,z} + \bH_{,\bz}) \dz \dbz
    - \left( z^2 \dbz^2 \bH_{,\bz} + \bz^2 \dz^2 H_{,z}\right)
    - \frac12\, (1 - z \bz) \left( \bz \dz H_{,z}  + z \dbz \bH_{,\bz} \right) \dot\Phi + \\ [3pt]
\disty \phantom{L = } \frac i4\, (1 + z \bz) \left( \bz \dz H_{,z}  - z \dbz \bH_{,\bz} \right) \left( \ima + \imb \right) + \\
\disty \phantom{L = i}  \frac 14\, \dot y e^{-\textstyle \frac{\Phi}2} \left(\rule{0pt}{1em} (1 + z \bz) (\ima + \imb)
    - 4i (z \dbz - \bz \dz )\right).
\ea
\ee
Now one may eliminate the auxiliary fields $\ima, \imb$ by their equations of motion and get desired Lagrangian with
four physical bosonic fields
\be\label{Lagr_fin}
\ba{l}
\disty L =  \M\, \frac{1 + z\bz}{1 - z\bz}\, \dz \dbz + \frac{\M^2 + \N^2}{16\M} (1 - z^2 \bz^2 )\dot{\tilde\Phi}^2
    + \frac{1 + z \bz }{(1 - z \bz)^3 } \frac{e^{-\tilde\Phi}}{\M}\,\dot y^2 -\\
\disty \phantom{L=-}  \frac12\, \bz (\M + i \N) \dot{\tilde\Phi} \dz -\frac12\, z (\M - i \N) \dot{\tilde\Phi} \dbz
    + 2i \frac{\bz \dz - z \dbz}{(1-z\bz)^2}\,e^{-\textstyle \frac{\tilde\Phi}2} \dot y
    - \frac 12 \frac{1 + z\bz}{1 - z\bz} \, \frac{\N}{\M} \, e^{-\textstyle \frac{\tilde\Phi}2} \, \dot{\tilde\Phi} \dot y.
\ea
\ee
where
\be\label{H_split}
\M = H_{,z} + \bH_{,\bz}, \qquad \N = -i (H_{,z} - \bH_{,\bz}).
\ee
and
\be\label{Phi_shift}
\tilde\Phi = \Phi - 2 \log(1 - z \bz).
\ee
One may check that this four-dimensional manifold is Ricci flat and, therefore, the sigma model (which possesses $N=8$
supersymmetry after restoration of all fermionic terms) is a hyper-K\"ahler one.

The action gets much simpler if one represents it in the Gibbons--Hawking form
\be\label{L}
S =  \int dt \left[ f \dot{\vec U}\,{}^2 + \frac1f \left( \dot y + \vec A \dot{\vec U} \right)^2\right],
\ee
with the triplet of coordinates $\vec U = \{u,\bar u, v \}$, where the new coordinates $u$, $\bar u$ and $v$ are defined as
\be\label{u}
u = \frac{z\, e^{-\tilde\Phi/2}}{1-z\bz}, \qquad \bar u =  \frac{\bz\, e^{-\tilde\Phi/2}}{1-z\bz},
\qquad v = \frac 12\,  e^{-\tilde\Phi/2}.
\ee
The function $f$ and vector $\vec A=\{A_u,A_{\bar u}, A_v \}$ are related to previously introduced quantities as
follows:
\bea
f(u,\bar u, v) &=& \M(z,\bz)\,\frac{(1-z\bz)^3}{1+z\bz} \,  \frac1{4v^2}, \nn
A_u (u,\bar u, v) &=&  i \,
\M(z,\bz)\,\frac{(1-z\bz)^3}{1+z\bz} \, \frac{\, \bar u }{8v^3},\nn
A_{\bar u} (u,\bar u, v) &=& i \,
\M(z,\bz)\, u
\,\frac{(1-z\bz)^3}{1+z\bz} \, \frac{\,  u }{8v^3},\nn
A_v (u,\bar u, v) &=& \N(z,\bz)\, (1-z\bz)^2\, \frac1{2v}\, .
\eea
Being written in terms of $u$, $\bar u$ and $v$ only (using (\ref{u})), the scalar $f$ and vector $\vec A$ satisfy the
conditions for the metric to be of the Gibbons--Hawking type \cite{GH}:
\be
\triangle f = 0, \qquad \vec\nabla \times \vec A =  \vec\nabla f.
\ee

Finally, it is instructive to present the simplest example of the Lagrangian \p{Lagr_fin} which corresponds to the choice
\be\label{free}
F(z)= \frac{z^2}{2}, \quad \bF(\bz)=\frac{\bz^2}{2}.
\ee
In this case the action is simplified drastically, and it reads
\be\label{Lfree}
S = -2\int dt  \left[ g \dot{\widehat\Phi}^2 + 16 \frac{\dx \dbx}g
    - 4 i \frac{\dot y }g \left( \bx \dx - x \dbx \right) e^{-\textstyle \frac{\widehat \Phi}2}
    + \frac14 \frac{\dot y}g^2 \, e^{- \widehat\Phi}
\right]
\ee
with the metric $g$ being
\be\label{metric}
g = \sqrt{\strut 1 - 4 x \bx }.
\ee
Here we have
\be\label{x}
x = \frac z{1 + z \bz}, \qquad \bx = \frac \bz {1 + z \bz}
\ee
and
\be\label{Phinew}
\widehat\Phi = \tilde \Phi +4  \log (1 + z \bz).
\ee
It is interesting to note that, with the choice \p{free}, the action for the LCM \p{finall} is a sum of free actions for
four bosonic fields, whereas in the case of the NCM we still have a non-trivial hyper-K\"ahler sigma model.

\section{Conclusion}
In the present paper we demonstrated that in $N=8$ supersymmetric mechanics with linear and nonlinear chiral
supermultiplets one may dualize two auxiliary fields into physical ones, in such a way that the bosonic manifold will
be a hyper-K\"ahler one. The key point of our construction is about different dualizations of the auxiliary components. One
auxiliary field is turned into a physical one in the standard way, through its replacement by the total time derivative of
some physical field. The other auxiliary field, which obeys the condition $\partial_t (\ima+\imb)=0$, is dualized through a
Lagrange multiplier. We clarified this choice of dualization by the illustrating the analogy with the three-dimensional
case.

The main difference of our procedure from the $N=4$ case \cite{KS} is its on-shell nature. We essentially used the equations
of motion and constraints for auxiliary components to turn them into physical ones. Our approach mimics the duality
transformations between a gauge vector field and scalar in $D=3$. The resulting supermultiplet is a $({\bf 4,8,4})$ one,
but it is still unclear whether it is on- or off-shell. So, the present result may be considered as a first,
preliminary one in a full understanding of $N=8, D=1$ supersymmetric sigma models with hyper-K\"ahler bosonic manifolds.
 In this respect it is desirable
to perform  a more detailed study of the dualization procedure in the case of $N=8, d=1$ supersymmetry. Of course, by
construction, the dualization will reproduce the restricted class of sigma models with properly constrained metrics of the
bosonic manifold (it depends on a smaller number of physical fields and, therefore, it possesses some additional isometries,
as compared to the general case). Nevertheless, such dualization goes in a much simpler way than the standard
construction of the most general sigma models. The main reason for this is that the unique way to have an interesting
geometry in the bosonic target space is to use nonlinear supermultiplets. Now it is not clear
how to systematically describe such supermultiplets in case of $N=8, d=1$ supersymmetry. Thus, the dualization may shed
light on the structure of such multiplets and the corresponding actions. In this respect it seems very interesting to
understand what happens after  dualization of one of the auxiliary component. The resulting off-shell
supermultiplet contains three physical bosons, while the set of auxiliary components is subjected to the differential
constraints. This is a new situation because the known treatments of a $({\bf 3,8,5})$ supermultiplet \cite{{ABC},{bikl}} did
not contemplate such possibility.

Another related question concerns the close analogy of our dualization procedure with (at least) three dimensional
theories with gauge vector fields. Moreover, the constraints on the auxiliary components we deal with in the cases of the
LCM and NCM look very similar to the situation with the four dimensional theory reduced to three dimensions, whereas the
dualization we used exhibits common features with the $c$-map \cite{SF}. It seems natural to expect the persistence of this
analogy along the line of the existence of new nonlinear supermultiplets in $D=3$, as well as to establish tighter
relations of our one-dimensional systems with their $D=4$ counterparts (radial motion in spherically symmetric
cases).

\section*{Acknowledgements}

This research was partially supported by the European Community's Marie Curie Research Training Network under contract
MRTN-CT-2004-005104 Forces Universe, and by the INTAS-00-00254 grant. S.K. thanks INFN --- Laboratori Nazionali
di Frascati  for the warm hospitality extended to them during the course of this work.

\end{document}